\documentclass[letterpaper, 10 pt, conference]{ieeeconf}  

\IEEEoverridecommandlockouts                              

\usepackage{graphicx}
\usepackage{amsmath} 
\usepackage{amssymb}  
\usepackage{dsfont}

\usepackage{xifthen}
\usepackage{xcolor}

\let\labelindent\relax
\usepackage{enumitem}

\newtheorem{theorem}{Theorem}[section]

\newtheorem{lemma}[theorem]{Lemma}
\newtheorem{proposition}{Proposition}
\newtheorem{corollary}{Corollary}
\newtheorem{assumption}{Assumption}
\newtheorem{definition}{Definition}
\newtheorem{example}{Example}
\newtheorem{remark}{Remark}

\newcommand{\sbar}{\Bar{s}}
\newcommand{\Abar}{\Bar{A}}
\newcommand{\Atilde}{\tilde{A}}
\newcommand{\Pbar}{\Bar{P}}
\newcommand{\Rset}{\mathbb{R}}
\newcommand{\Eset}{\mathrm{E}}
\newcommand{\Id}{\mathbb{I}}
\newcommand{\Ncal}{\mathcal{N}}
\newcommand{\Scal}{\mathcal{S}}
\newcommand{\Fcal}{\mathcal{F}}
\newcommand{\Gcal}{\mathcal{G}}
\newcommand{\gradJ}{\nabla_{s_i}J_i}
\newcommand{\smax}{s_{\rm max}}
\renewcommand{\P}{\mathbb{P}}

\title{\LARGE \bf
Gradient Dynamics in Linear Quadratic Network Games with Time-Varying Connectivity and Population Fluctuation}

\author{Feras Al Taha, Kiran Rokade, Francesca Parise 
\thanks{This work was supported in part by the Natural Sciences and Engineering Research Council of Canada.}
\thanks{The authors are with the School of Electrical and Computer Engineering, Cornell University, Ithaca, NY, 14853, USA (e-mail:  \{\tt\small\{foa6,kvr36,fp264\}@cornell.edu)}%
}

\begin{document}

\maketitle
\thispagestyle{empty}
\pagestyle{empty}

\begin{abstract}

In this paper, we consider a learning problem among non-cooperative agents interacting in a time-varying system. Specifically, we focus on repeated linear quadratic network games, in which the network of interactions changes with time and agents may not be present at each iteration. To get tractability, we assume that at each iteration, the network of interactions is sampled from an underlying random network model and agents participate at random with a given probability. Under these assumptions, we consider a gradient-based learning algorithm and establish almost sure convergence of the agents’ strategies to the Nash equilibrium of the game played over the expected network. Additionally, we prove, in the large population regime, that the learned strategy is an $\epsilon$-Nash equilibrium for each stage game with high probability. We validate our results over an online market application.

\end{abstract}

\section{INTRODUCTION}

The prominence of big data analytics, Internet-of-Things, and machine learning has transformed large societal and technological systems.
Of special interest are systems in which a large number of independent agents interacting over a network structure make strategic decisions, while affected by the actions of other agents.
Examples include pricing in demand-side electricity markets, supply chain management, wireless networks, public goods provision, financial exchanges, and international trades, among others.
In many settings, agents lack information or computational capabilities to determine a priori the best strategy to guide their decisions.
Instead, agents need to learn their best action over time based on observations of their rewards and other agents actions, while adapting to changes in their environment.

Standard models of learning dynamics assume that the population of agents and their underlying network of interactions are static.
However, in systems with large populations and high-volume interactions, this assumption is not realistic since individuals may  enter and leave the system, leading to fluctuations in the population size, and participating agents might change their interconnections with other agents over time.
It then becomes unclear whether players can reach a stable outcome in this nonstationary environment.

Accordingly, there is a need for a theory of multi-agent learning in dynamic network settings in which the composition of the agents and their interactions can  change over time.
In this paper, we start addressing these issues by focusing on gradient play dynamics in linear quadratic network games with time-varying connectivity and dynamic population.

\subsection{Main contributions}

To obtain a tractable framework, we model the time-varying network of interactions and population of agents as stochastic network realizations from an underlying known distribution.
Using stochastic approximation methods \cite{jiang2008stochastic}, we demonstrate that for linear quadratic network games, when all agents follow a projected gradient descent scheme, they almost surely converge to a Nash equilibrium of the game played over the expected network.
Moreover,  by using concentration inequalities,  we show that with high probability, the learned strategy profile is an $\epsilon$-Nash equilibrium of the game played over any realized network, where $\epsilon$ decreases as the population size increases.

\subsection{Related works}

\textit{Learning in time-varying settings with dynamic populations} has been previously studied for games with a special structure such as congestion games, bandwidth allocation, markets, first-price auctions or public good games \cite{lykouris2016learning,bergemann2010dynamic,cavallo2009efficient,cavaliere2012prosperity}.
The efficiency of outcomes in such games was investigated for low-adaptive-regret learning \cite{lykouris2016learning} and later generalized to low-approximate-regret learning \cite{foster2016learning}, under the assumption that the population size is fixed.
The setting with a changing number of agents was studied for congestion games in \cite{shah2010dynamics}. Similarly, the effect of changing populations has been studied in the context of truthful mechanism design  \cite{dolan1978incentive,parkes2003mdp, cavallo2009efficient}. None of these works cover the setting of network games considered in this paper.

In terms of learning dynamics to reach Nash equilibria in \textit{static noncooperative games or multi-agent settings}, many schemes have been studied.
These include fictitious play \cite{fudenberg1998theory}, parallel and distributed computation \cite{bertsekas2015parallel}, projection-based algorithms (e.g., projected gradient dynamics) \cite{facchinei2003finite}, and regret minimization (e.g., no-regret dynamics) \cite{nisan2007algorithmic}.
Among these, a growing literature considered learning dynamics for \textit{static aggregative and network games} focusing, for example, on best response dynamics \cite{parise2019variational,parise2020distributed}, projection-based algorithms \cite{paccagnan2016aggregative,salehisadaghiani2018distributed}, forward--backward operator splitting  \cite{gadjov2020single,grammatico2017proximal}, passivity-based schemes \cite{gadjov2018passivity,de2019distributed}, distributed asynchronous schemes \cite{koshal2016distributed,zhu2022asynchronous}, and state-dependent costs \cite{shokri2021network}, among others.

When considering non-stationary environments, the setting of \textit{agents communicating according to a time-varying network} is not novel in the control literature.
Specifically, a large literature focused on cooperative problems such as consensus/gossip algorithms \cite{moreau2005stability,hendrickx2012convergence} as well as distributed optimization \cite{nedic2014distributed} over time-varying networks, mainly based on an assumption of uniform connectivity over contiguous intervals of time.
Similar results have been derived for distributed Nash equilibrium seeking in noncooperative games \cite{grammatico2017proximal,belgioioso2020distributed,cenedese2020asynchronous,salehisadaghiani2016distributed} and for averaging algorithms over randomized networks with gossip communication \cite{fagnani2008randomized,boyd2006randomized}.
One common property of these works is that despite the changing communication network, the underlying problem to solve (achieving consensus, optimizing some objective in a decentralized fashion, or reaching a Nash equilibrium) is time-invariant. 
In contrast, in this paper, agents take part in a different network game at every iteration. Thus, the Nash equilibrium that the players are trying to learn is itself time-varying since their payoff function depends on the time-varying network. 
The question at hand is then not solely about the convergence of  agents' strategies, but also about the nature of the learned strategy and its relation to the time-varying stage games. 
We show that, in our setting, strategies converge to the equilibrium of the game played over the expected network and that this is approximately optimal for large populations.
Finally, convergence of optimistic gradient descent in games with time-varying utility functions has recently been studied in 
\cite{anagnostides2023convergence}. 
Our model is different as it applies to network games in which payoff-variability is due to variability of interconnections. 
Moreover, we  consider a setting in which the population itself may vary, with agents joining and leaving the system.
Open multi-agent systems with random arrivals and departures of agents have been recently studied for cooperating homogeneous agents in consensus dynamics \cite{hendrickx2017open} or optimal resource allocation problems \cite{de2022random}. 
We instead focus on non-cooperating agents with network interactions.

\subsection{Paper organization}

The rest of the paper is organized as follows.
Section~\ref{sec:recap} presents the repeated network game setup and recaps known results on learning dynamics in static environments.
Section~\ref{sec:dyn_net} introduces the main results on the convergence of gradient dynamics and the suboptimality guarantees of the learned strategy for a time-varying network and fixed population.
Section \ref{sec:dyn_pop} extends these results to dynamic populations.
Section \ref{sec:num_exp} demonstrates the convergence of the proposed projected gradient dynamics with numerical simulations and Section \ref{sec:concl} concludes the paper.
Omitted proofs are given in the Appendix. 

\subsection{Notation}
We denote by $[v]_j$ the $j$th component of a vector $v\in\Rset^n$ and by $A_{ij}$ the $ij$th entry of a matrix $A\in\Rset^{m\times n}$. 
We denote the Frobenius norm of a matrix by $\|\cdot\|_F$ and the Euclidean norm of a vector as well as the matrix norm induced by the Euclidean norm by $\|\cdot\|_2$.
The symbol $\Id_n$ denotes the $n\times n$ identity matrix.
The operator $\Pi_\mathcal{X}[\cdot]$ denotes the projection onto the set $\mathcal{X}$.
The symbol $\otimes$ denotes the Kronecker product. 
Given $n$ matrices $A_1, \dots, A_n$ of appropriate dimensions, we let $\textrm{diag}(A_1,\dots,A_n)$ denote the matrix formed by stacking $A_1, \dots, A_n$ block-diagonally.
Given vectors $x_1,\dots,x_n$, we let $[x_i]_{i\in\{1,\dots,n\}}$ denote the vector formed by stacking $x_1,\dots,x_n$ vertically.
For a given square matrix $A$, $\lambda_{\rm max}(A)$ and $\lambda_{\rm min}(A)$ denote the maximum and minimum eigenvalues of $A$, respectively.

\section{RECAP ON LINEAR QUADRATIC GAMES OVER STATIC NETWORKS} \label{sec:recap}

We present a recap of known results for static games.

\subsection{One shot game} \label{sec:lq_game}

Consider a linear quadratic (LQ) game played by a population of agents indexed by $i\in\Ncal:=\{1,\dots,N\}$ over a static network with no self-loops, represented by an adjacency matrix $\Atilde \in [0,1]^{N\times N}$.
Each agent $i\in\Ncal$ aims at selecting a strategy $s_i  \in \Scal_i \subseteq \Rset^n$ to minimize a cost function
\begin{align} \label{eq:lq_cost}
    J_i(s_i, s_{-i}, \Atilde) = \frac{1}{2} s_i^\top Q_i s_i -  s_i^\top \Big (\theta_i  + \frac{\alpha}{N} \sum_{j=1}^N \Atilde_{ij} s_j \Big )
\end{align} 
where $s_{-i} := [s_j]_{j\in\Ncal \setminus \{ i\}} \in \Rset^{(N-1)n}$ is the strategy profile of all other agents, $\alpha\in\Rset$ models the strength of network effects, $Q_i\in\Rset^{n\times n}$ and $\theta_i \in \Rset^n$ are parameters modeling agent specific heterogeneity, and $Q_i=Q_i^\top \succ 0$. Let $\mathcal{S} := \prod_{i \in \Ncal} \Scal_i$ be the set of all strategy profiles of the game.
Given a strategy profile $s:= [s_i]_{i\in\Ncal} \in \Scal$ and a network $\Atilde$, the term $ \frac{1}{N} \sum_{j=1}^N \Atilde_{ij} s_j$ represents the local aggregate sensed by agent $i$.
We denote this LQ game by $\Gcal(Q,\theta,\alpha,\Atilde)$ where $\theta:=[\theta_i]_{i\in\Ncal}$ and $Q:= \textrm{diag}(Q_1,\dots,Q_N)$.

\begin{definition}[$\epsilon$-Nash equilibrium]
Given $\epsilon>0$, a strategy profile $\sbar \in \Rset^{N n}$ is an $\epsilon$-Nash equilibrium of the game $\Gcal(Q,\theta,\alpha,\Atilde)$ if for all $i\in\Ncal$, we have $\sbar_i \in \Scal_i$ and
\begin{align} \label{eq:eps_NE}
    J_i(\sbar_i, \sbar_{-i}, \Atilde) \le J_i(s_i, \sbar_{-i}, \Atilde) + \epsilon \qquad \textrm{for all } s_i \in \Scal_i.
\end{align}
If \eqref{eq:eps_NE} holds for $\epsilon=0$, then $\sbar$ is a Nash equilibrium.
\end{definition}

It is well known that the Nash equilibria of convex games can be characterized by using the game Jacobian
\begin{align}
    F(s, \Atilde) &:= [\gradJ(s_i,s_{-i},\Atilde)]_{i\in\Ncal}
\end{align}
which is a map composed of each player's cost gradient with respect to their own strategy.
For LQ games, this is
\begin{align}
    F(s, \Atilde) 
    &= \Big [Q_i s_i - \theta_i -\frac{\alpha}{N} \sum_{j=1}^N \Atilde_{ij} s_j \Big]_{i\in\Ncal} \nonumber\\
    &= Q s - \theta -\frac{\alpha}{N} (\Atilde \otimes \Id_n) s \nonumber.
\end{align}
A classic approach to characterizing Nash equilibria is to interpret them as solutions to variational inequalities
(see \cite{facchinei2003finite,scutari2010convex} for general games and \cite{parise2019variational} for network games).
We summarize this relation in the following lemma, after introducing an assumption that guarantees existence and uniqueness of the Nash equilibrium.
\begin{assumption}[Equilibrium uniqueness] \label{a:eq_uniq}
\begin{enumerate}[label=(\roman*)]
    \item []
    \item For each $i\in\Ncal$, $\Scal_i$ is nonempty, convex and compact. There exists a compact set $\mathcal{\bar{S}} \subseteq \Rset^n$ such that $\Scal_i\subseteq \mathcal{\bar{S}}$ for all $i$, and $s_{\rm max}:= \max_{s\in\mathcal{\bar{S}}} \|s\|_2 <\infty$. \label{a:strat_set}
    \item \label{a:unique} The following relation holds
    \begin{align} 
        \lambda_{\rm min}(Q) - \frac{|\alpha|}{N} \|\tilde{A}\|_2  > 0.
    \end{align} 
\end{enumerate}
\end{assumption}

\begin{lemma}{(Variational inequality equivalence, \cite[Proposition~1.4.2]{facchinei2003finite}, \cite[Proposition~1 and Proposition~2]{parise2019variational})} \label{lem:VI}
    Suppose that Assumption \ref{a:eq_uniq}-\ref{a:strat_set} holds. 
    Then, the strategy profile $\sbar$ is a Nash equilibrium of $\Gcal(Q,\theta,\alpha,\Atilde)$ if and only if it solves the variational inequality $(s-\sbar)^\top F(\sbar,\Atilde) \ge 0, \quad \forall s \in \mathcal{S}.$ If, additionally, Assumption~\ref{a:eq_uniq}-\ref{a:unique} holds, then, $F$ is strongly monotone and the equilibrium is unique.
\end{lemma}

\subsection{Learning dynamics in repeated games}

Consider now a repetition of the stage LQ game defined in Section \ref{sec:lq_game}, played over the static network $\Atilde$.
Suppose that each agent $i$ tries to learn the equilibrium iteratively by using the projected gradient dynamics
\begin{align} \label{eq:grad_dyn}
    s^{k+1} = \Pi_{\mathcal{S}} [ s^k - \tau  F(s^k, \Atilde) ],
\end{align}
where $s^k$ is the strategy profile at iteration $k\ge 0$, $s^0 \in \Scal$, and $\tau>0$ is a fixed step size. 

\begin{remark} 
    To implement their learning dynamics
    \begin{align*} 
        \textstyle s^{k+1}_i = \Pi_{\Scal_i} \left[ s_i^k - \tau  \left(Q_i s_i^k - \theta_i -\frac{\alpha}{N} \sum_{j=1}^N \Tilde{A}_{ij} s_j^k\right) \right],
    \end{align*}
    player $i\in\Ncal$ only requires knowledge of their current strategy, their cost function parameters, and their current local aggregate. The player does not need to observe the network realization $\Tilde{A}$.
\end{remark}

The dynamics in \eqref{eq:grad_dyn} can be shown to converge under a mild assumption on the step size.

\begin{lemma}{(Gradient play over a static network, \cite[Theorem~12.1.2]{facchinei2003finite})} \label{lem:static_dyn} 
    Suppose that Assumption \ref{a:eq_uniq} holds. 
    Let $L:= \lambda_{\rm max}(Q) + \frac{|\alpha|}{N}\|\Atilde\|_2$ be the Lipschitz constant of $F$ and $\mu :=\lambda_{\rm min}(Q) - \frac{\alpha}{N} \| \Atilde\|_2$ be the strong monotonicity constant of $F$.
    If the step size $\tau$ satisfies $0<\tau< 2\mu/L^2$, then for any initial strategy profile $s^0\in\Scal$, the gradient dynamics described in \eqref{eq:grad_dyn} converge to the unique Nash equilibrium of the game $\Gcal(Q,\theta,\alpha,\Atilde)$.
\end{lemma}
The Lipschitz and strong monotonicity constants used in Lemma \ref{lem:static_dyn} are computed, e.g., in \cite{parise2019variational}.

\section{GAMES OVER TIME-VARYING NETWORKS} \label{sec:dyn_net}

The well-known results summarized in the previous section hold for a static network of interactions.
In many realistic settings however, interactions among the agents may change across repetitions of the game and agents may not always be participating.
In this section, we consider learning dynamics in time-varying networks.
These results are then extended in Section \ref{sec:dyn_pop} to the case of dynamic populations.

More specifically, we here assume that the network of interactions is random, and at every repetition $k = 0, 1, 2, \dots$ of the stage game, a new independent network realization $A^k \in [0,1]^{N \times N}$ is sampled as follows:
for all $i\in\Ncal$, $A_{ii}^k=0$ (no self-loop) and for $i\ne j$, $A_{ij}^k$ is sampled from a distribution with bounded support\footnote{Restricting the support to the unit interval is without loss of generality. One can easily extend this paper's results to any bounded convex support.} in $[0,1]$ and with mean $\Abar_{ij}\in[0,1]$. 
In the case where $A_{ij}^k$ is a Bernoulli random variable, $\Abar_{ij}$ can be interpreted as the probability that player $j$ contributes to the local aggregate of player $i$. 
More generally, $\Abar=\Eset[A^k]$ represents the expected network of interactions, where we set $\Abar_{ii}=0$ for all $i\in\Ncal$.

\subsection{Motivating example}
We motivate this time-varying setting by presenting an example in the context of online markets.

\begin{example}[Dynamic pricing game] \label{ex:mkt}
Consider an online market with $N$ sellers. 
Each merchant $i \in \Ncal$ sells a product~$i$ at a price $s_i \in \Rset_+$ determined daily.
On each day $k = 0, 1, 2, \dots$, a new set of $M$ customers arrives. Each customer $c \in \{1,\dots,M\}$ decides which products to purchase. 
The demand of customer $c$ on day $k$ for product $i$ can be modeled as an affine demand function \cite{yue2006bertrand}
\begin{align}
    d_i^{c,k} = \Bar{d}_i - \eta \Big (s_i^k + \frac{\alpha}{N} \sum_{j=1}^N A_{ij}^{c,k} s_j^k \Big)
\end{align} 
where $s^k_i$ is the price of product $i$ on day $k$, $\bar{d}_i$ is the maximum demand that a customer can have for product $i$, $\eta > 0$ is the price sensitivity, $\alpha\ge 0$ models the degree of influence that the price of other products has on the demand of product $i$, and $A_{ij}^{c,k} \sim \textrm{Ber}(\Bar{A}_{ij})$ is a Bernoulli random variable that indicates whether customer $c$ is interested in co-purchasing product $i$ with product $j$ on day $k$. 
The mean $\Bar{A}_{ij} \in [0,1]$ can be interpreted as the likelihood of this event. 

The total demand for product $i$ can be obtained by aggregating the demands of all customers:
\begin{align*}
    d_i^k &= \sum_{c=1}^M d_i^{c,k} = \sum_{c=1}^M \Big(\Bar{d}_i - \eta \Big (s_i^k + \frac{\alpha}{N}\sum_{j=1}^N A_{ij}^{c,k} s_j^k \Big)\Big)\\
    &= M \Bar{d}_i -  \eta \Big ( M s_i^k + \frac{\alpha}{N} \sum_{j=1}^N \sum_{c=1}^M  A_{ij}^{c,k} s_j^k \Big)\\
    &= M \Big( \bar{d}_i -  \eta \Big( s_i^k + \frac{\alpha}{N} \sum_{j=1}^N A^k_{ij} s_j^k \Big)\Big)
\end{align*}
where $A^k_{ij} := (1/M)\sum_{c=1}^M A_{ij}^{c,k}$  is a random variable representing the average complementarity of products $i$ and $j$, with mean $\Eset[A^k_{ij}]=\Abar_{ij}$.
Given this demand, each merchant tries to maximize their profit, i.e., minimize the cost
\begin{align} \label{eq:mkt_ex_cost}
    J_i(s_i^k, s_{-i}^k, &A^k) = - s_i^k d_i^k \nonumber \\
    &= -  s_i^k M \Big ( \Bar{d}_i - \eta \Big (s_i^k + \frac{\alpha}{N}\sum_{j=1}^N A^k_{ij} s_j^k\Big ) \Big ) \nonumber \\
    &=  M \Big(\eta (s_i^k)^2 - s_i^k \Big(\bar{d}_i - \eta \frac{\alpha }{N}\sum_{j=1}^N A^k_{ij} s_j^k\Big)\Big).
\end{align}
This online market competition can be interpreted as a repeated LQ game on a dynamic network $A^k$ that changes daily to capture different daily customer preferences. 
Sellers may learn how to price their products using a gradient descent scheme.
Note that the gradient in this case is
\begin{align}
\label{eq:grad_mkt}
    \gradJ (s_i^k,s_{-i}^k,A^k) = -d_i^k + \eta M s_i^k,
\end{align}
hence, the only information seller $i$ requires to compute their current gradient update is the total number of customers and the quantity of product $i$ sold on the previous day. 
\end{example}

In the next section, we characterize the asymptotic behavior produced by gradient updates in time-varying settings, as motivated by the previous example, for general LQ games.

\subsection{Learning dynamics}

In this nonstationary environment, the Nash equilibrium is time-varying.
Despite this, we show that agents learn to play a suitable fixed strategy, provided that they use a time-varying, diminishing step size $\tau^k$ to ensure convergence of their gradient-based learning dynamics
\begin{align} \label{eq:grad_for_dyn_net}
    s^{k+1} = \Pi_{\Scal} [ s^k - \tau^k  F(s^k, A^k) ]
\end{align}
where $k \geq 0$, and $s^0\in\Scal$. 
The diminishing step size is needed to compensate for the network variability.

\begin{assumption}[Step size] \label{a:step_size}
    For all $k$, the step size $\tau^k$ satisfies $\tau^k>0$, $\tau^k \to 0$, $\sum_{k=1}^\infty \tau^k = \infty$ and $\sum_{k=1}^\infty (\tau^k)^2 < \infty$.
\end{assumption}

In the following, we show that the gradient dynamics \eqref{eq:grad_for_dyn_net} converge almost surely to the Nash equilibrium of the LQ game played over the expected network $\Abar = \Eset[A^k]$. 
This result follows from rewriting the dynamics as a stochastic gradient descent scheme and utilizing stochastic approximation theory \cite{jiang2008stochastic} to show convergence.

\begin{proposition} \label{prop:dyn_net}
Suppose that Assumptions \ref{a:eq_uniq} (for $\Atilde=\Abar$) and \ref{a:step_size} hold. 
Then, the gradient dynamics in \eqref{eq:grad_for_dyn_net} converge to the unique Nash equilibrium of the game played over the expected network $\Abar$ almost surely, for any $s^0\in\Scal$. 
\end{proposition}

If agents compute their gradient updates with noisy local aggregate data, convergence can still be achieved provided that the noise is additive, zero-mean, and with bounded variance (as it can be absorbed in the perturbation vector~$w^k$).

The following corollary justifies why the Nash equilibrium of the game played over the expected network is a reasonable policy to learn, by showing that with high probability, it results in an approximate equilibrium for any iteration.

\begin{corollary} \label{cor:epsilon_Nash} 
Suppose that Assumption \ref{a:eq_uniq} (for $\Tilde{A} = \Bar{A}$) holds. Fix any $k \geq 0$ and any $\delta>0$. Then, with probability at least $1-\delta$, the Nash equilibrium of the LQ game $\Gcal(Q,\theta,\alpha,\Abar)$ played over the expected network $\bar{A}$ is an $\epsilon_{N,\delta}$-Nash equilibrium to the stage game $\Gcal(Q,\theta,\alpha,A^k)$ played over the network $A^k$ with     
\begin{align*}
    \epsilon_{N,\delta} := 2 |\alpha| \smax^2 \sqrt{\frac{n\ln\left(2nN/\delta\right)}{2N}}.\\[-5pt] 
\end{align*}
\end{corollary}

We note that $\epsilon_{N,\delta}$ does not depend on the cost parameters $Q_i$ and $\theta_i$. 
Also, for any confidence $\delta>0$, $\epsilon_{N,\delta}$ decreases as the population size $N$ grows. 
This corollary can be proven by using matrix concentration inequalities, similarly to the approach in \cite[Lemma 14]{parise2023graphon}.

\section{GAMES OVER DYNAMIC POPULATION} \label{sec:dyn_pop}

In this section, we generalize the previous setting by considering a dynamic population, where players randomly enter and leave the game from one repetition to the other.
For example, in the online market setting introduced in Example~\ref{ex:mkt}, sellers might lack supply for their products or decide to not sell their items on certain days.
This can be modeled in the following way.
If a player does not participate in a particular iteration, then their strategy does not update in that iteration, resulting in dynamics where a subset of components are updated at a time, similarly to random coordinate descent dynamics.

Let $P^k\in \{0,1\}^{N\times N}$ be a diagonal matrix such that the random variable $P_{ii}^k\sim\textrm{Ber}(\bar{P}_{ii})$ represents whether player~$i$ is participating ($P_{ii}^k=1$) in the game at iteration $k$ or not ($P^k_{ii}=0$), with $\Pbar_{ii}>0$ for all $i\in\Ncal$. 
Then, the network realization over which the game is played at time $k$ is $A^k P^k$ (when $P_{ii}^k=0$, the corresponding column in the adjacency matrix is set to zero such that player $i$ does not affect any other players participating in the game). 

\noindent Consider the following gradient dynamics for each player $i$ 
\begin{align} \label{eq:grad_for_dyn_pop}
    s^{k+1}_i\!&=\!\begin{cases}
    \Pi_{\mathcal{S}_i} \Big[ s^k_i - \dfrac{\tau^k}{\Pbar_{ii}} \nabla_{s_i}J_i(s_i^k,s^k_{-i}, A^k P^k) \Big] & \text{if}\, P_{ii}^k=1,\\
    s_i^k & \text{if}\, P_{ii}^k=0,
    \end{cases}
\end{align}
where $ k \geq 0$, and $s_i^0 \in \Scal_i$. 
Note that in \eqref{eq:grad_for_dyn_pop}, each player $i$ scales their gradient step by $1/\Pbar_{ii}$ to compensate for missing updates when $P_{ii}^k=0$.

Despite these random interruptions to the sequence of strategy updates, we show that the players learn to play the Nash equilibrium of the game $\Gcal(Q,\theta,\alpha,\Abar\Pbar)$ played over the expected network $\bar{A}\bar{P}$. 
Similar to Proposition \ref{prop:dyn_net}, the key step to prove this result is to rewrite the dynamics \eqref{eq:grad_for_dyn_pop} as stochastic gradient dynamics converging to the aforementioned Nash equilibrium.

\begin{proposition} \label{prop:dyn_pop}
Suppose that Assumptions \ref{a:eq_uniq} (with $\tilde{A}=\Abar\Pbar$) and \ref{a:step_size} hold.
If $\|\theta_i\| \le \theta_{\rm max}$ for all $i\in\Ncal$, then, the gradient dynamics described in \eqref{eq:grad_for_dyn_pop} converge to the unique Nash equilibrium of the LQ game played over the expected network $\Bar{A}\bar{P}$ almost surely, for any $s^0 \in \Scal$. 
\end{proposition}

Similar to Corollary \ref{cor:epsilon_Nash}, the Nash equilibrium of the expected network can be shown to be an approximate Nash equilibrium to the stage game with high probability.

\begin{corollary}
\label{cor:epsilon_Nash_dynamic_pop}
    Suppose that Assumption \ref{a:eq_uniq} (for $\Tilde{A} = \Bar{A}$) holds. Fix any $k \geq 0$ and $\delta>0$. Then, with probability at least $1-\delta$, the Nash equilibrium of the LQ game played over the network $\bar{A} \bar{P}$ is an $\epsilon_{N,\delta}$-Nash equilibrium to the stage game played over the network $A^k P^k$, with 
    \begin{align*}
        \epsilon_{N,\delta} := 2 |\alpha| \smax^2 \sqrt{\frac{n\ln\left(2nN/\delta\right)}{2N}}.\\[-10pt]
    \end{align*}
\end{corollary}

\section{NUMERICAL EXPERIMENTS} \label{sec:num_exp}

In this section, we present numerical experiments demonstrating the convergence of gradient dynamics for the dynamic pricing game presented in Example \ref{ex:mkt} for both the static and the dynamic population settings presented in Sections \ref{sec:dyn_net} and \ref{sec:dyn_pop}, respectively. 

Consider a market with $N$ sellers, where each seller is endowed with a product whose demand has price sensitivity $\eta=1$.
Let $\alpha=0.8$ be the strength of network effects on the demand of every product.
We consider two categories of products, namely category $1$ and category $2$. 
The probability that a customer is interested in co-purchasing products from categories $m$ and $l$ is denoted by $\bar{a}_{ml}$, where $m,l\in\{1,2\}$. 
Thus, in the notation of Example \ref{ex:mkt}, $\Bar{A}_{ij} = \bar{a}_{ml}$ if product $i$ is in category $m$ and product $j$ is in category $l$. 
For the numerical simulations in this section, we set these probabilities to be $\Bar{a}_{11} = \Bar{a}_{22} = 0.8$ and $\Bar{a}_{12} = \bar{a}_{21} = 0.3 $. 
Moreover, let $\Bar{d}_1 = 2$, $\Bar{d}_2 = 10$ be the maximum demands a customer can have for products in categories $1$ and $2$, respectively. 

\begin{figure}
    \centering
    \includegraphics[width=\columnwidth]{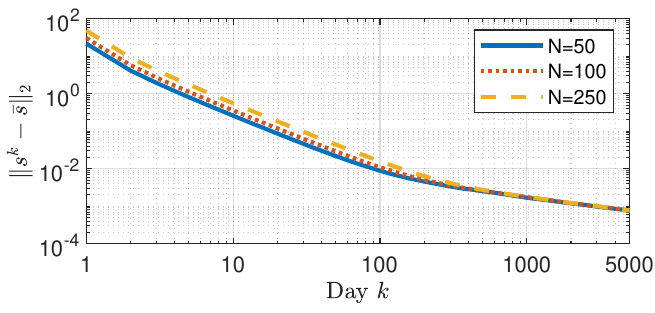}
    \includegraphics[width=\columnwidth]{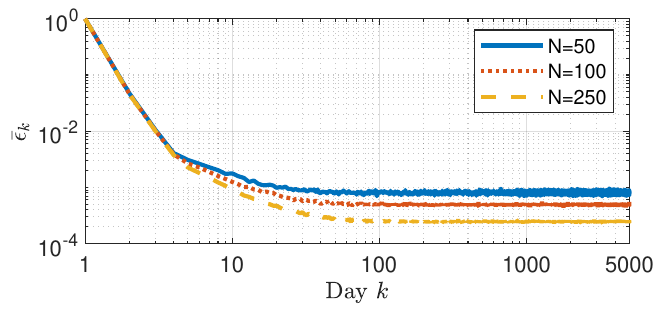}
    \caption{For dynamic network $A^k$ and fixed population size $N$. Top: Convergence of product prices $s^k$ (averaged over 100 trials) to the Nash equilibrium price profile $\sbar$ of the game played over the expected network $\Bar{A}$. Bottom: Maximum suboptimality gap (averaged over 100 trials) across sellers per iteration as defined in \eqref{eq:epsilon_k}. 
    }
    \label{fig:dyn_net}
\end{figure}

We first consider the case of static population, in which all sellers participate in the market every day. 
Each seller starts with an arbitrary initial price for their product (set to zero in our simulations). 
On each day $k$, a new set of $M=100$ customers arrives. 
The total demand $d_i^k$ for product $i$ on day $k$ is determined by the realization of the co-purchasing preference matrix $A^k$ sampled according to the matrix $\Bar{A}$. 
With the gradient of their cost function computed as in \eqref{eq:grad_mkt}, seller $i$ updates the price of their product following the projected gradient dynamics in \eqref{eq:grad_for_dyn_net} with step size $\tau_k = 1/(Lk)$. 
For various values of $N$, we illustrate in Fig. \ref{fig:dyn_net} (top) the deviation of the resulting sequence of price profiles $\{s^k\}_{k\ge 0}$ from the Nash equilibrium $\Bar{s}$ of the game played over the expected co-purchasing matrix $\Bar{A}$, averaged over $100$ trials. 
It can be observed that the prices converge to the price profile $\sbar$ for each $N$, as predicted in Proposition~\ref{prop:dyn_net}.

\begin{figure}
    \centering
    \includegraphics[width=\columnwidth]{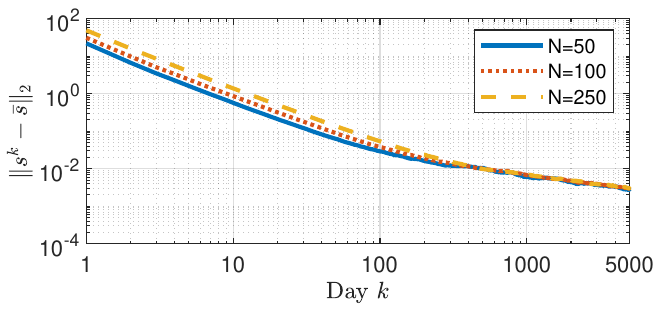}
    \includegraphics[width=\columnwidth]{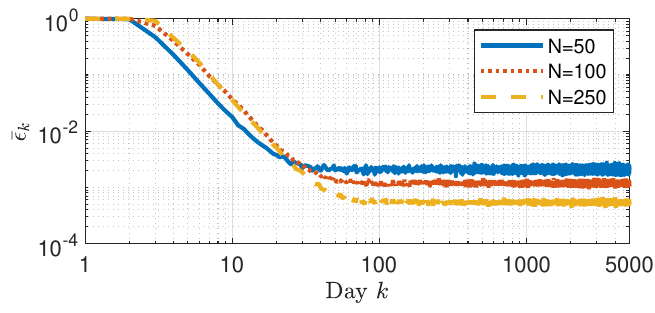}
    \caption{For dynamic network $A^kP^k$ and dynamic population. Top: Convergence of product prices $s^k$ (averaged over 100 trials) to the Nash equilibrium price profile $\sbar$ of the game played over the expected network $\Bar{A}\Bar{P}$. Bottom: Maximum suboptimality gap (averaged over 100 trials) across sellers per iteration as defined in \eqref{eq:epsilon_k} with $A^kP^k$ instead of $A^k$.}
    \label{fig:dyn_pop}
\end{figure}

For each day $k$, we compute the largest \textit{normalized suboptimality gap} incurred across sellers defined as
\begin{align}
\label{eq:epsilon_k}
    \bar{\epsilon}_k := \max_{i\in\Ncal} \left|\frac{J_i(s_i^k,s_{-i}^k,A^k) - J_i^\star(A^k)}{J_i^\star(A^k)}\right|
\end{align}
where $J_i^\star(A^k) := \inf_{s \in\Scal_i} J_i(s,s^k_{-i},A^k)$ is the cost incurred by seller $i$ by best-responding to other sellers' prices. 
Fig.~\ref{fig:dyn_net} (bottom) depicts the decrease in the gap $\bar{\epsilon}_k$ for different population sizes as a function of $k$.
The oscillations that can be observed in Fig. \ref{fig:dyn_net} (bottom) are due to the variability of the network, and are smaller for larger population sizes.

Next, we consider the dynamic population case, where some sellers may not participate in the market every single day. 
Let $\bar{p} = 0.9$ be the probability that any seller $i\in\Ncal$ participates in the market on a given day. Thus, $\bar{P} = \Bar{p} \ \Id_N$. 
In this setting, the sellers use the gradient dynamics in \eqref{eq:grad_for_dyn_pop} with a step size of $\tau_k = 1/(Lk)$. 
The deviation between the obtained sequence of prices $\{s^k\}_{k\ge 0}$ from the Nash equilibrium $\Bar{s}$ of the game played over the expected co-purchasing matrix $\Bar{A} \Bar{P}$ is shown in Fig. \ref{fig:dyn_pop} (top) for various values of $N$, averaged over 100 trials. 
The slower convergence of the price profile relative to the static population case is expected since sellers randomly skip updating their prices on certain days.
Similar to the static population case, Fig. \ref{fig:dyn_pop} (bottom) presents the largest suboptimality gap across sellers for each iteration $k$, averaged across 100 trials.
A similar trend to Fig. \ref{fig:dyn_net} (bottom) can be observed with the decrease of $\bar{\epsilon}_k$ as a function of $N$, but with larger oscillations.
The latter is also to be expected given that the network $A^k P^k$ has more variability than the network $A^k$.

\section{CONCLUSIONS} \label{sec:concl}

In this work, we have introduced a tractable framework for learning in network games with time-varying connectivity and dynamic populations.
Several extensions and future research directions can be considered. 
For example, in this paper, we have restricted our attention to myopic learning dynamics. 
However, one could define a discounted cumulative/averaged payoff over all iterations and examine the outcome of forward-looking dynamics. 
Other interesting directions would be to extend the model to generic cost functions and evolving graphs that densify with time \cite{leskovec2005graphs}.

\bibliographystyle{ieeetr}
\bibliography{references}

\section{APPENDIX}

\subsection{Proof of Proposition \ref{prop:dyn_net}}
The gradient dynamics in \eqref{eq:grad_for_dyn_net} can be interpreted as stochastic gradient play for the LQ game played over the expected network $\Abar$.
This can be seen by rewriting the dynamics as 
\begin{align*}
    s^{k+1} &= \Pi_{\Scal} [ s^k - \tau^k F(s^k, A^k) ]\\
    &= \Pi_{\Scal} [ s^k - \tau^k ( F(s^k, \Abar) + \underbrace{F(s^k, A^k) - F(s^k, \Abar)}_{=:w^k} ) ]
\end{align*}
where $w^k$ can be interpreted as a stochastic perturbation vector. 

Let $\{\Fcal_k\}_{k\geq 0}$ denote an increasing sequence of $\sigma$-algebras such that $s^k$ is $\Fcal_k$-measurable. 
Consider the following conditions on the perturbation vector $w^k$ and the operator $F$:
\begin{enumerate}[label=(\roman*)]
    \item The perturbation has zero mean $\Eset[w^k \mid \Fcal_k] = 0$; 
    \item The variance of $w^k$ is such that $\sum_{k=1}^\infty (\tau^k)^2 \Eset[\|w^k\|^2_2 \mid \Fcal_k] < \infty$  almost surely;
    \item The operator $F(\cdot,\Abar)$ is $L$-Lipschitz;
    \item The operator $F(\cdot,\Abar)$ is $\mu$-strongly monotone.
\end{enumerate}
By Assumption \ref{a:step_size} and \cite[Theorem 3.2]{jiang2008stochastic}, it suffices to show that the conditions (i)-(iv) hold to guarantee that the stochastic projected gradient dynamics presented above converge to the solution of the variational inequality in Lemma~\ref{lem:VI} with network~$\Abar$.
Then, by Lemma \ref{lem:VI}, this solution is a Nash equilibrium of the game with network $\bar{A}$. 

Conditions (iii) and (iv) are satisfied with constants defined in Lemma \ref{lem:static_dyn}.
\begingroup
\allowdisplaybreaks
To show conditions (i) and (ii), we first compute a simpler expression for the perturbation vector $w^k$
\begin{align*}
    w^k &= F(s^k,A^k) - F(s^k,\Bar{A}) \\
    &= \left ( Q s^k - \theta  -\frac{\alpha}{N} (A^k \otimes \Id_n) s^k \right )\\
    &\qquad - \left( Q s^k -  \theta  - \frac{\alpha}{N} (\Abar \otimes \Id_n) s^k \right)\\
    &= - \frac{\alpha}{N} ((A^k - \bar{A}) \otimes \mathbb{I}_n) s^k.
\end{align*}
Condition (i) holds since 
\begin{align*}
    \Eset[w^k \mid \Fcal_k ] &=  \Eset\left [- \frac{\alpha}{N} ((A^k - \bar{A}) \otimes \mathbb{I}_n) s^k  \mid \Fcal_k  \right ]\\
    &= - \frac{\alpha}{N} ((\Eset[A^k] - \bar{A}) \otimes \mathbb{I}_n) s^k   =0.
\end{align*}
Condition (ii) also holds since 
\begin{align*}
    \Eset&[\| w^k\|_2^2 \mid \Fcal_k ] = \Eset \Big [ \left \| - \frac{\alpha}{N} ((A^k - \bar{A}) \otimes \mathbb{I}) s^k  \right\|_2^2 \ \Big \vert \ \Fcal_k \Big]\\
    &\le  \Eset \Big [ \frac{\alpha^2}{N^2} \|(A^k - \bar{A}) \otimes \mathbb{I}\|_2^2 \|s^k \|_2^2 \ \Big \vert \  \Fcal_k \Big]\\
    &\overset{(a)}{\le}  \Eset \Big [ \frac{\alpha^2}{N^2} \|A^k - \bar{A} \|_F^2 N \smax^2  \Big]\\
    &\overset{(b)}{\le}   \frac{\alpha^2}{N^2} N^3 \smax^2 
    = N \alpha^2 \smax^2 < \infty
\end{align*}
where (a) follows from the fact that the 2-norm is upper bounded by the Frobenius norm and (b) follows from $|A_{ij}^k - \Abar_{ij}| \leq 1$ since both terms take value in $[0,1]$. 
By Assumption~\ref{a:step_size}, $\Eset[\, \|w^k\|^2_2 \, \vert \, \Fcal_k]\le N \alpha^2 s_{\rm max}^2 <\infty$ implies that $\sum_{k=1}^\infty (\tau^k)^2 \Eset[\, \|w^k\|^2_2 \,\vert\, \Fcal_k] $ is  upper bounded by the convergent series $  N\alpha^2 s_{\rm max}^2 \sum_{k=1}^\infty (\tau^k)^2  $ and is thus convergent.
\endgroup

Therefore, by \cite[Theorem 3.2]{jiang2008stochastic}, the stochastic gradient dynamics converge to the unique solution of the variational inequality $(s-\sbar)^\top F(\sbar,\Abar) \ge 0, \quad \forall s \in \Scal$ almost surely and the desired result follows from Lemma~\ref{lem:VI}. 

\subsection{Proof of Corollary \ref{cor:epsilon_Nash}}
    The proof follows the method of \cite[Lemma~14]{parise2023graphon}. 
    Let $\sbar$ be the Nash equilibrium of the game played over the network~$\bar{A}$.
    For any player $i \in \Ncal$,
    \begingroup
    \allowdisplaybreaks 
    \begin{align}
        &J_i(\sbar_i, \sbar_{-i}, A^k) - \inf_{s_i\in\Scal_i} J_i(s_i,\sbar_{-i}, A^k) \nonumber \\
        &\quad = \frac{1}{2} \sbar_i^\top Q_i \sbar_i - \sbar_i^\top \Big ( \theta_i + \frac{\alpha}{N}\sum_{j=1}^N A^k_{ij} \sbar_j\Big ) - \inf_{s_i\in\Scal_i} \Big\{\frac{1}{2} s_i^\top Q_i s_i \nonumber \\
        &\qquad - s_i^\top \Big ( \theta_i + \frac{\alpha}{N}\sum_{j=1}^N (A^k_{ij}+\bar{A}_{ij}-\bar{A}_{ij}) \sbar_j \Big )\Big\} \nonumber \\
        &\quad \le \frac{1}{2} \sbar_i^\top Q_i \sbar_i - \sbar_i^\top \Big ( \theta_i + \frac{\alpha}{N}\sum_{j=1}^N A^k_{ij} \sbar_j\Big ) \nonumber \\
        &\qquad - \inf_{s_i\in\Scal_i} \Big\{\frac{1}{2} s_i^\top Q_i s_i - s_i^\top \Big ( \theta_i + \frac{\alpha}{N}\sum_{j=1}^N \bar{A}_{ij} \sbar_j \Big )\Big\} \nonumber \\
        &\qquad - \inf_{s_i\in\Scal_i} -\frac{\alpha}{N} s_i^\top  \sum_{j=1}^N (A^k_{ij}-\bar{A}_{ij}) \sbar_j  \nonumber \\
        &\quad \overset{(a)}{=} \frac{\alpha}{N} \sbar_i^\top  \sum_{j=1}^N (\bar{A}_{ij}-A^k_{ij}) \sbar_j+ \sup_{s_i\in\Scal_i} \frac{\alpha}{N} s_i^\top  \sum_{j=1}^N (A^k_{ij}-\bar{A}_{ij}) \sbar_j \nonumber \\
        \label{eq:2_alpha_over_N}
        &\quad \overset{(b)}{\leq} 2 \frac{|\alpha|}{N} s_{\rm max} \Big\|\sum_{j=1}^N (\bar{A}_{ij}-A^k_{ij}) \sbar_j\Big\|_2
    \end{align}
    \endgroup
    where (a) holds since $\sbar_i$ is the minimizer of the first infimum term, and (b) follows from the Cauchy-Schwarz inequality. 
    
    Next, we derive a high-probability bound on $\left\|\sum_{j=1}^N (\bar{A}_{ij}-A^k_{ij}) \sbar_j\right\|_2$. 
    For all $j \in\Ncal$, for all $l \in \{1, \dots, n\}$, let $(\bar{A}_{ij} - A^k_{ij}) [\bar{s}_{j}]_l =: X^j_l$ and $\sum_{j =1}^N X^j_l =: X_l$ (the $j=i$ term is simply zero). Then,
    \begin{align*}
        \Big\|\sum_{j =1}^N (\bar{A}_{ij}-A^k_{ij}) \sbar_j\Big\|^2_2 &=  \sum_{l = 1}^n \Big(\sum_{j =1}^N (\bar{A}_{ij} - A^k_{ij}) [\bar{s}_{j}]_l\Big)^2 \\
        &= \sum_{l = 1}^n (X_l)^2. 
    \end{align*}
    Note that for each $l$, the random variables $\{X_l^j\}_{j \in \Ncal}$ are independent since $\{A^k_{ij}\}_{j \in \Ncal}$ are independent; $\Eset[X^j_l] = \Eset\left[(\bar{A}_{ij} - A^k_{ij}) [\bar{s}_{j}]_l\right] = 0$ since $\Eset[A^k_{ij}] = \Bar{A}_{ij}$; and 
    $|X^j_l|= |(\bar{A}_{ij} - A^k_{ij}) [\bar{s}_{j}]_l| = |\bar{A}_{ij} - A^k_{ij}|\cdot |[\bar{s}_{j}]_l| \leq \smax$ since $|\Abar_{ij}-A^k_{ij}| \leq 1$ due to the fact that $A_{ij}^k,\Abar_{ij}\in [0,1]$. 
    Hence, by Hoeffding's inequality, for all $t > 0$,
    \begin{align*}
        \P\left(|X_l| \geq t\right) \leq 2 \exp\left(\frac{-2t^2}{N \smax^2}\right). 
    \end{align*} 
    To upper bound $\P\left(|X_l| \geq t\right)$ by $\delta/nN$, we fix $t $ such that
    \begin{align*}
        &2 \exp\left(\frac{-2t^2}{N \smax^2}\right) = \frac{\delta}{nN} \\
        \iff &t^2 = \frac{N \smax^2}{2} \ln\left(\frac{2nN}{\delta}\right). 
    \end{align*}
    By the union bound, $\P\left(|X_l| \leq t, \, \forall l=1,\dots,n \right) \ge 1- \delta/N$.
    Hence, with probability at least $1 - \delta/N$, $\sum_{l = 1}^n X_l^2 \leq n t^2$
    and thus
    \begin{align*}
        &\Big\|\sum_{j =1}^N (\bar{A}_{ij}-A^k_{ij}) \sbar_j\Big\|_2 \leq \sqrt{n} t
        &\leq \sqrt{\frac{nN\smax^2\ln\left(2nN/\delta\right)}{2}} . 
    \end{align*}

    Substituting the bound obtained above into \eqref{eq:2_alpha_over_N}, we obtain that for any arbitrary player $i \in \Ncal$, with probability at least $1 - \delta/N$,
    \begin{align*}
        &J_i(\sbar_i, \sbar_{-i}, A^k) - \inf_{s_i\in\Scal_i} J_i(s_i,\sbar_{-i}, A^k) \\
        &\leq 2 |\alpha| \smax^2 \sqrt{\frac{n\ln\left(2nN/\delta\right)}{2N}}\\ 
        &=: \epsilon_{N,\delta}. 
    \end{align*}
    By the union bound, we obtain that, with probability at least $1 - \delta$,
    \begin{align*}
        J_i(\sbar_i, \sbar_{-i}, A^k) - \inf_{s_i\in\Scal_i} J_i(s_i,\sbar_{-i}, A^k)
        \leq \epsilon_{N,\delta}
    \end{align*}
    for all players $i \in \Ncal$.
    Thus, with probability at least $1 - \delta$, $\Bar{s}$ is an $\epsilon_{N,\delta}$-Nash equilibrium of the game $\Gcal(Q,\theta,\alpha,A^k)$ played over $A^k$.     

\subsection{Proof of Proposition \ref{prop:dyn_pop}}
We first rewrite the gradient dynamics in \eqref{eq:grad_for_dyn_pop} to interpret them as stochastic gradient play. 
Consider the strategy update of player $i$ 
\begingroup
\allowdisplaybreaks
\begin{align*}
    &s^{k+1}_i = \Pi_{\mathcal{S}_i} \Big[ s^k_i - \frac{\tau^k}{\Pbar_{ii}} P^k_{ii}  \nabla_{s_i} J_i(s^k_i,s^k_{-i},A^k P^k) \Big] \\
    &= \Pi_{\mathcal{S}_i} \Big [ s^k_i - \frac{\tau^k}{\Pbar_{ii}} P^k_{ii}  \Big (Q_i s_i^k -\theta_i-\frac{\alpha}{N}\sum_{j=1}^N A_{ij}^k P_{jj}^k s_j^k \Big) \Big ] \\
    &= \Pi_{\mathcal{S}_i} \Big [ s^k_i - \frac{\tau^k}{\Pbar_{ii}}   \Big (P^k_{ii} Q_i s_i^k + \bar{P}_{ii} Q_i s_i^k - \bar{P}_{ii} Q_i s_i^k - P^k_{ii} \theta_i\\
    &\qquad \qquad   + \bar{P}_{ii} \theta_i -\bar{P}_{ii} \theta_i -\frac{\alpha }{N}\sum_{j=1}^N P^k_{ii} A_{ij}^k P_{jj}^k s_j^k  \Big ) \Big] \\
    &= \Pi_{\mathcal{S}_i} \Big [ s^k_i - \frac{\tau^k}{\Pbar_{ii}}   \Big ( \bar{P}_{ii} Q_i s_i^k - \bar{P}_{ii} \theta_i  -\frac{\alpha }{N}\sum_{j=1}^N P^k_{ii} A_{ij}^k P_{jj}^k s_j^k\\
    &\qquad \qquad  + (P_{ii}^k- \bar{P}_{ii} ) (Q_i s_i^k - \theta_i) \Big ) \Big ] \\
    &= \Pi_{\mathcal{S}_i} \Big [ s^k_i - \tau^k  \Big  (  Q_i s_i^k -  \theta_i -\frac{\alpha }{N} \sum_{j=1}^N \Eset[\Tilde{A}_{ij}^k] s_j^k +w_i^k \Big) \Big] 
\end{align*}
\endgroup
where $w_i^k := \frac{\alpha }{N} \sum_{j=1}^N (\Eset[\Tilde{A}_{ij}^k] - \tilde{A}_{ij}^k ) s_j^k + \frac{1}{\Pbar_{ii}}(P_{ii}^k- \bar{P}_{ii} ) (Q_i s_i^k - \theta_i) $ and $\Tilde{A}_{ij}^k := \frac{1}{\Pbar_{ii}}P^k_{ii} A_{ij}^k P_{jj}^k$. 
Note that $$\Eset[\tilde{A}^k]=\Eset[\Pbar^{-1} P^k A^k P^k]=\bar{A}\bar{P}.$$ 
This is because $\Eset[\tilde{A}^k_{ij}]= \frac{1}{\Pbar_{ii}} \Eset[P^k_{ii} A^k_{ij} P^k_{jj}]=\frac{1}{\Pbar_{ii}}\bar{P}_{ii}\bar{A}_{ij}\bar{P}_{jj}$ if $i\neq j$ since the random variables are independent, and $\Eset[\tilde{A}^k_{ii}]=\frac{1}{\Pbar_{ii}}\Eset[P^k_{ii} A^k_{ii} P^k_{ii}]=0$ since $A^k_{ii}=0$. 
Hence, we obtain the following stochastic gradient dynamics for all agents
\begin{align*}
    s^{k+1} &= \Pi_{\mathcal{S}} [s^k - \tau^k (F(s^k,\Abar \Pbar )  + w^k ) ]
\end{align*}
where $w^k = [w_i^k]_{i\in\Ncal}$ is the perturbation vector.
The dynamics of this game can be interpreted as a stochastic approximation to the gradient play over the game with expected network $\Abar\Pbar$. 

As in Proposition \ref{prop:dyn_net}'s proof, we can verify conditions (i)-(iv) to show convergence of these stochastic projected gradient dynamics. 
One can show that conditions (i)-(ii) hold with the same approach used in Proposition \ref{prop:dyn_net} by showing that $\Eset[\, w^k \, \vert \, \Fcal_k]=0$ and that $\Eset[\, \|w^k\|_2^2 \, \vert \, \Fcal_k] < \infty$.
Conditions (iii)-(iv) also hold, by Lipschitzness and strong monotonicity of $F(\cdot, \Abar\Pbar)$.

Therefore, by \cite[Theorem 3.2]{jiang2008stochastic}, the stochastic gradient dynamics will converge almost surely to the unique solution of the variational inequality $(s-\Bar{s})^\top F(\sbar,\Abar\Pbar) \ge 0$, for all $s \in \Scal$ and the desired result follows from Lemma \ref{lem:VI}.

\subsection{Proof of Corollary \ref{cor:epsilon_Nash_dynamic_pop}}
    The proof is similar to that of Corollary \ref{cor:epsilon_Nash} and the same arguments hold. 
    Note that for a fixed $i$, the random variables $ \{  A^k_{ij} P^k_{jj} \}_{j \in \Ncal}$ have the same properties as their counterparts $\{A^k_{ij}\}_{j\in\Ncal}$ in Corollary~\ref{cor:epsilon_Nash}. 
    In particular, the random variables $\{ \bar{A}_{ij} \bar{P}_{jj} - A^k_{ij} P^k_{jj} \}_{j \in \Ncal}$ are independent, zero-mean and satisfy $| \bar{A}_{ij}\bar{P}_{jj} - A^k_{ij}P^k_{jj} | \leq1$, and $ A^k_{ii} P^k_{ii} = 0$ since $A^k_{ii} = 0$. 


\end{document}